\documentclass[aip,jmp,amssymb,superscriptaddress,footinbib]{revtex4-1} 

\usepackage{amssymb}
\usepackage{graphicx}
\usepackage{subfigure}
\usepackage{color}
\usepackage{hyperref}
\hypersetup{
 colorlinks,
linkcolor=red,
filecolor=green,
urlcolor=blue,
citecolor=blue
}

\def\ket#1{|#1\rangle }

\begin{document}

\title{Effects of three-body collisions in a two-mode Bose-Einstein condensate}
%\address{Institute for Theoretical Physics, Technical University of Berlin,
%Hardenbergstr. 36, D-10623, Berlin, Germany}
\author{Carlos Sab{\'\i}n}
\address{School of Mathematical Sciences, University of Nottingham, Nottingham NG7 2RD
United Kingdom}
\author{Pablo Barberis-Blostein}
\address{Instituto de Investigaciones en Matematicas Applicadas y Sistemas, Universidad Nacional Aut\'onoma de M\'exico}
\author{Cristopher Hern\'andez}
\address{Instituto de F\'isica, Universidad Nacional Aut\'onoma de M\'exico}
\author{Robert B. Mann}
\address{Perimeter Institute for Theoretical Physics, 31 Caroline St N, Waterloo, Ontario, N2L 2Y5, Canada}\address{Department of Physics \& Astronomy, University of Waterloo, Waterloo, Ontario, N2L 3G1, Canada}
\author{Ivette Fuentes}
\address{School of Mathematical Sciences, University of Nottingham, Nottingham NG7 2RD
United Kingdom}

\date{\today}

\begin{abstract}
We study the effects of three-body collisions in the physical
properties of a two-mode Bose-Einstein condensate. The model introduced here includes two-body and
three-body elastic and mode-exchange collisions and can be solved analytically. We will use this fact to show
that three-body interactions can produce drastic changes in the probability distribution of the ground state and the dynamics
of the relative population. In particular, we find that three-body
interactions under certain circumstances may inhibit the collapse of the
relative population.
\end{abstract}
%\pacs{03.75.Gg, 42.50.Gy,03.75.Lm} 
\maketitle

\section{Introduction}
Most of our understanding of condensed matter is based on models
which consider two-body collisions.  However, there are many situations
where three-body or higher order collisions are relevant in the physical
properties of such systems \cite{three, phases, spinliquids}. For
example, it is known that three-body collisions are important
in systems that show exotic quantum
phases, such as topological phases \cite{phases} or spin liquids
\cite{spinliquids}. Moreover, it is suspected that many-body collisions
are important in the coldest phases of Bose-Einstein condensates,
where the dilute regime breaks down \cite{dilute}. 
Microscopic calculations show that polar molecules driven by
microwave fields undergo three-body interactions \cite{zoller}.
The interaction potentials of molecules trapped in an optical
lattice give rise to Hubbard models with strong nearest-neighbour
two-body and three-body interactions. 

In this letter we find the
exact analytical solution of a generalized two-mode Bose-Hubbard
model which includes two-body and three-body elastic and
mode-exchange collisions. Then we show that three-body collisions can 
change dramatically the properties of the ground state of a two-mode
Bose-Einstein condensate. The effects are also observable in the
evolution of the relative population inhibiting, in some cases,
quantum collapse.  It is well known that three-body collisions
are responsible for particle loss in Bose-Einstein condensates, through a process
called three-body recombination \cite{3bodyrecomb}. During three-body collisions the
particles recombine to form a molecule which is not trapped by the
potential. However, it is now possible to
inhibit molecular three-body recombination in atomic Bose-Einstein
condensates via the application of resonant $2\pi$ laser pulses
\cite{search}. In such situations our model becomes of special
interest, since it takes into account three-body collisions where particles do
not recombine and remain trapped in the potential.

The model we introduce can describe the physics of a
double-well Bose-Einstein condensate or a spin-1/2 Bose-Einstein
condensate consisting of particles with two internal degrees of
freedom trapped in a single well. In the context of the double-well
Bose-Einstein condensate, the mode-exchange collisions included here are known as
generalized nearest neighbour interactions \cite{heiselberg} and give
rise to coherent tunneling effects \cite{tunneling,bloch}. Recent
analysis show that stronger two-body interactions are correlated
with two-body coherent tunneling dynamics in which two particles
simultaneously tunnel through the barrier \cite{tunneling}. This
effect, also known as second order tunneling, has been
observed in the laboratory \cite{bloch}. Mode-exchange collisions
are called inelastic collisions in the context of spin-$1/2$
condensates and occur when cold collisions take place in the
presence of light fields. Such is the case of spin-$1/2$ condensates
where a laser field is used to induce Josephson-type interactions,
which produce transitions among the spin degrees of freedom \cite{julienne}.

\begin{figure}[t!]
\begin{center}
\includegraphics[width=0.4\textwidth]{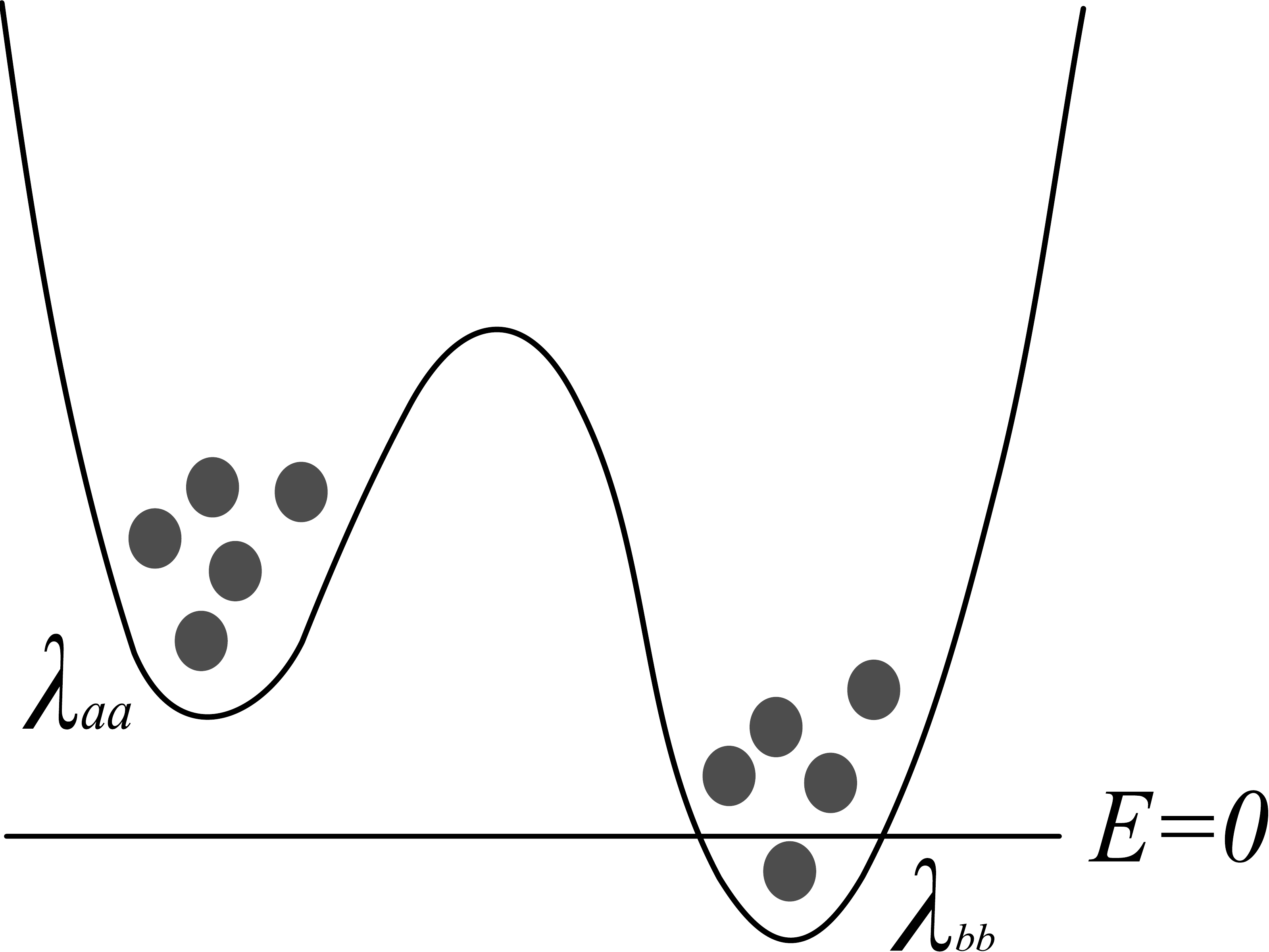}
\caption{A Bose-Einstein condensate in an asymmetric double-well potential, characterised by the single well energies $\lambda_{a|a}$ and $\lambda_{b|b}$.}
\label{double}
\end{center}
\end{figure}

\section{Model}

We consider a general model of a two-mode Bose-Einstein condensate that
includes two-body and three-body collisions given by the Hamiltonian
$\mathcal{H}_3=H_{1}+H_{2}+H_{3}$ where (taking $\hbar=1$):
\begin{eqnarray}\label{hamiltonian3}
H_{1}&=&\lambda_{a|a}a^\dagger a+\lambda_{b|b}b^\dagger b+ \lambda_{a|b}(a^\dagger b+b^\dagger a),\nonumber\\
H_{2}&=&\mathcal{U}_{aa|aa}a^\dagger a^\dagger
aa+\mathcal{U}_{bb|bb}b^\dagger b^\dagger b
b+\mathcal{U}_{ab|ab}a^\dagger
b^\dagger ab\nonumber\\&+& U_{aa|ab}(a^\dagger a^\dagger ab+h.c.)+U_{bb|ab}(b^\dagger b^\dagger ab+h.c.)\nonumber\\
&+&U_{aa|bb}(a^\dagger a^\dagger bb+ h.c.),\nonumber\\
H_{3}&=&\mathcal{U}_{aaa|aaa}a^\dagger a^\dagger a^\dagger
aaa+\mathcal{U}_{bbb|bbb}b^\dagger b^\dagger b^\dagger
bbb \nonumber\\& &\mathcal{U}_{aab|aab}a^\dagger a^\dagger b^\dagger
aab+\mathcal{U}_{abb|abb}a^\dagger b^\dagger b^\dagger
abb \nonumber\\& +&U_{aaa|aab}\,(a^\dagger a^\dagger a^\dagger
aab+h.c.)\nonumber\\&+&U_{bbb|abb}\,(b^\dagger b^\dagger b^\dagger
abb+h.c.) \nonumber\\&+ &U_{aaa|abb} \,(a^\dagger a^\dagger a^\dagger
abb+h.c.)\nonumber\\ &+&U_{bbb|aab}(b^\dagger b^\dagger b^\dagger
aab+h.c.) \nonumber\\&+&U_{aab|abb}(a^\dagger a^\dagger b^\dagger
abb+h.c.)\nonumber\\&+&U_{aaa|bbb}(a^\dagger a^\dagger a^\dagger
bbb+h.c.).
\end{eqnarray}

The operators $a^{\dagger},a$ and $b^{\dagger},b$ are associated
with two modes, labeled A and B, with respective frequencies $\lambda_{a|a}$
and $\lambda_{b|b}$. These modes correspond either to atoms with two
different hyperfine levels \cite{JILA2}, or two spatially separated
condensates \cite{MIT} (see Fig.(\ref{double})). The Josephson-type interaction, in
which one particle is annihilated in one mode and created in the
other, has coupling constant $\lambda_{a|b}$. This process
 is induced by applying a magnetic field gradient 
\cite{MIT} or a laser \cite{JILA2}. The terms in $H_{2}$,
which have four bosonic operators, describe two-particle collisions.
The two-body elastic scattering strengths are given by
$\mathcal{U}_{aa|aa}$ and $U_{bb|bb}$ for same mode collisions and
$U_{ab|ab}$  when the particles colliding belong to different modes.
Mode-exchange collisions have interaction strengths
$U_{aa|ab},U_{bb|ab}$ when two particles collide and one of them is
transformed into the other mode and interaction strength $U_{aa|bb}$
when the collision transforms two particles in one mode into the
other mode. This process is also know as second order tunneling in
the context of a double-well BEC \cite{tunneling,bloch}. 

The Hamiltonian
$\mathcal{H}_{2}=H_{1}+H_{2}$ has been studied in detail in
\cite{fuent2}. This two-body interaction Hamiltonian coincides with
the two-mode Bose-Hubbard model if mode-exchange
collisions are neglected. i.e. $U_{aa|ab}=U_{bb|ab}=U_{aa|bb}=0$.
However, microscopical calculations show that such interactions,
known as inelastic collisions in the context of spin-1/2
Bose-Einstein condensates, should be considered since they occur
when particles collide in the presence of a laser field
\cite{julienne}. Surprisingly, including such collision allows for an
exact analytical solution \cite{fuent1,fuent2,fuent3,fuent4}. Here we include a
three-body collision term given by $H_{3}$, where three-body
interactions consist of products of six operators (three creation
and three annihilation). This term includes all possible three-body
collisions where $\mathcal{U}_{aaa|aaa}$, $\mathcal{U}_{bbb|bbb}$,
$\mathcal{U}_{aab|aab}$ and $\mathcal{U}_{abb|abb}$ correspond to elastic
scattering lengths and
$\mathcal{U}_{aaa|aab}$, $\mathcal{U}_{aaa|abb}$,
$\mathcal{U}_{aaa|bbb}$ correspond to mode-exchange collisions where
one, two and three particles change mode, respectively.

We have found that the Hamiltonian $\mathcal{H}_{3}$ has six
families of exact analytical solutions. In this paper we present
the solution which we consider of greatest physical interest. The
other solutions will be presented elsewhere.

We start by considering the double-well potential
shown in Fig.(\ref{double}). Particles undergo two- and three- body collisions
and we assume that first, second and third order tunneling events
can occur. In second (third) order tunneling two (three) tunneling
events can occur coherently. Therefore single particles can
coherently tunnel two (three) times and two (three) particles can
tunnel simultaneously during a collision. 

We consider that a particle in well A (or
B) has probability amplitude $A_1\cos\theta$ (or $-A_1\cos\theta$)
of staying in well A (or B) and probability amplitude
$A_1\sin\theta$ of tunneling to well B (or A). $A_1$ is the first
order tunneling strength and $\theta$ is the tunneling phase. Note
that the minus sign appears because we chose for simplicity well B
to have negative energy corresponding to
$\lambda_{b|b}=-\lambda_{a|a}$. We consider $A_2$ and $A_3$ to be
second and third order tunneling strengths.
Therefore $A_2\sin^2\theta$ and $A_3\sin^2\theta\cos\theta$ for example, are
the second and third order probability amplitudes respectively, for
a single particle in well A to tunnel back and forth. 

The
coefficients in the single particle Hamiltonian $H_1$ are found by
considering all possible single particle events including second and
third order tunneling. For example,
\begin{eqnarray}\label{eq:lambdaa}
\lambda_{a|a}&=&A_1\cos\theta+A_2(\cos^2\theta+\sin^2\theta)\\\nonumber &+&A_3\cos\theta(\cos^2\theta+\sin^2\theta)=A_2+ (A_3+A_1)\cos{\theta},
\end{eqnarray}
is the probability amplitude for a single particle in well A to end
in well A. The general two-body and three-body scattering lengths
$U_{ij|lm}$ and $U_{ijk|lmn}$ are given by the product of the
corresponding second and third order tunneling strengths times the
appropriate tunneling phase amplitudes ($\sin\theta$ if the
particle tunnels during the collision and  $\pm\cos\theta$ if the particle stays). For
example, consider a three-body collision during which two particles
change state. The total probability amplitude will be
 \begin{equation}\label{eq:uaaaabb}
 U_{aaa|abb}=3A_3\cos\theta\sin^2\theta. 
 \end{equation}
 The factor 3 comes from
the fact that there are three possible events that give rise to the
same final outcome, according to the different time ordering of the events. 

In the case of two-body collisions we consider
 that during a collision two and three tunneling events can
occur. So collisions in which two particles in well A end up both
in well B is given by
\begin{eqnarray}\label{eq:uaabb}
U_{aa|bb}&=&A_2\sin^2\theta\\\nonumber&+&3A_3(\cos\theta\sin^2\theta-\sin^2\theta\cos\theta)=A_2\sin^2\theta,
\end{eqnarray}
where again the factor 3 comes from the time ordering. Such considerations  give rise to the parameters,
\begin{eqnarray}\label{eq:coef}
  \lambda_{a|a}&=&A_2+(A_{3}+A_1)\cos\theta, \nonumber\\
\lambda_{b|b} &=&A_2-(A_{3}+A_1)\cos\theta,\nonumber\\
 \lambda_{a|b} &=&(A_{1}+A_{3})\sin\theta,\nonumber\\
\mathcal{U}_{aa|aa}&=&(A_{2}\cos\theta+3A_{3})\cos\theta,\nonumber\\
\mathcal{U}_{bb|bb}&=& (A_{2}\cos\theta-3A_{3})\cos\theta,\nonumber\\
  \mathcal{U}_{ab|ab}&=&2A_{2}(\sin^2\theta-\cos^2\theta), \nonumber\\
  \mathcal{U}_{aa|ab} &=&(3A_{3}+2A_{2}\cos\theta)\sin\theta ,\nonumber\\
  \mathcal{U}_{bb|ab}&=&(3A_{3}-2A_{2}\cos\theta)\sin\theta,\nonumber\\
\mathcal{U}_{aa|bb} &=& A_2\sin^2\theta,\nonumber\\
\mathcal{U}_{bbb|bbb}&=&-\mathcal{U}_{aaa|aaa}=-A_{3}\cos^3\theta,\nonumber\\
\mathcal{U}_{abb|abb}&=&-\mathcal{U}_{aab|aab}=-A_{3}(2\cos\theta\sin^2\theta-\cos^3\theta),\nonumber\\
\mathcal{U}_{aaa|aab}&=&\mathcal{U}_{bbb|abb}=3A_{3}\cos^2\theta\sin\theta ,\nonumber\\
\mathcal{U}_{aaa|abb}&=&3A_3\cos\theta\sin^2\theta,\nonumber\\
\mathcal{U}_{bbb|aab}&=&-3A_3\cos\theta\sin^2\theta,\nonumber\\
\mathcal{U}_{aab|abb}&=&3A_{3}(\sin^3\theta-\cos^2\theta\sin\theta),\nonumber\\
\mathcal{U}_{aaa|bbb}&=&A_{3}\sin^3\theta.
\end{eqnarray}

At this point it is important to illustrate the connection of this model with a physical model of a BEC. 
 Consider the many-body energy functional for bosonic particles of mass $m$ trapped in a potential $V(\mathbf{r})$ undergoing two-body  and three-body collisions:
\begin{eqnarray} \label{eq:hamiltonian}
H&=& H_0+ H_{I2}+H_{I3},\nonumber\\
 H_0&=&\int d\mathbf{r} \big(-\frac{\hbar^2}{2m}\hat\Psi^{\dagger}\nabla^2\hat\Psi+\hat\Psi^{\dagger}V(\mathbf{r})\hat\Psi\big),\nonumber\\
&=&\int d\mathbf{r} \hat\Psi^{\dagger}H_t\hat\Psi,\nonumber\\
H_{I2}&=&\frac{g_2}{2}\int d\mathbf{r}\hat\Psi^{\dagger}\hat\Psi^{\dagger}\hat\Psi\hat\Psi,\nonumber\\
H_{I3}&=&\frac{g_3}{2}\int d\mathbf{r}\hat\Psi^{\dagger}\hat\Psi^{\dagger}\hat\Psi^{\dagger}\hat\Psi\hat\Psi\hat\Psi,
\end{eqnarray}
where  $g_2, g_3$ are two-body and three-body coupling strengths respectively and $H_t$ is the  Hamiltonian of the trap.  The wavefunction $\hat\Psi$ can be expanded in terms of  a certain set of functions $\phi_i$  and their corresponding annihilation operators $\hat c_i$  as 
\begin{equation}\label{eq:modes}
\hat\Psi=\sum_i\phi_i\hat c_i. 
\end{equation}
We employ the standard two-mode approximation
\begin{equation}\label{eq:2modes}
\hat\Psi=\phi_1\hat c_1+\phi_2\hat c_2 
\end{equation}
and then the rotation
\begin{eqnarray}\label{eq:lmodes}
\phi_1&=&\cos({\theta}/{2})\phi_a-\sin({\theta}/{2})\phi_b\nonumber\\
\phi_2&=&\cos({\theta}/{2})\phi_b+\sin({\theta}/{2})\phi_a,
\end{eqnarray}
where $\phi_a$, $\phi_b$ are nearly-normalized modes  \cite{milburn} with  $\int dr\phi_a\phi_a=1+\epsilon$, $\int dr\phi_b\phi_b=1-\epsilon$, where the amplitude of transition between them 
\begin{equation}\label{eq:epsilon}
\epsilon=\int\,dr\,\phi_aH_t\phi_b 
\end{equation}
is assumed to be very small. We obtain the Hamiltonian $\mathcal{H}_3$ described above with:
\begin{eqnarray}\label{trapasimmetry}
A_1=\frac{1}{2}(E_1-E_2);\, A_2=\frac{1}{2}U_2;\,U_2=U_{11}=U_{22};\, A_3=\frac{1}{2}U_3;\,U_3=U_{111}=U_{222}\nonumber\\
E_{1}=\int dr\phi_1H_t\phi_1\, ,\,  U_{11}=\int dr\phi_1^4, \,  U_{111}=\int dr\phi_1^6,\nonumber\\
E_{2}=\int dr\phi_2H_t\phi_2\, , \, U _{22}=\int dr\phi_2^6,\,  U_{222}=\int dr\phi_2^6,
\end{eqnarray}
plus several terms of order $\epsilon$, $\epsilon^2$ and $\epsilon^3$ which can be treated perturbatively \cite{fuent3,fuent4} as long as $\epsilon << A_1\theta$. We will discuss the experimental validity of this approximation in the last section. 

\section{Results}
Surprisingly, the Hamiltonian $\mathcal{H}_3=H_{1}+H_{2}+H_{3}$ has an exact analytical solution for
this set of parameters. The analytical expression of its eigenstates is:
\begin{equation}\label{eq:eigenstates}
\ket{E_m}=e^{\frac{-\theta}{2}(a^{\dagger}b-ab^{\dagger})}\ket{J,m}, 
\end{equation}
where $|J,m\rangle$ are the Fock states for $2J=N$. $N$ is the total number of particles, 
given by the operator:
\begin{equation}\label{eq:number}
\hat{N}=n_{a}+n_{b}=a^{\dagger}a+b^{\dagger}b ,
\end{equation}
and $m$ is the
eigenvalue of the relative population operator:
\begin{equation}\label{eq:unitary}
\hat{m}=(a^{\dagger}a-b^{\dagger}b)/2 . 
\end{equation}
Since the number of particles in the system $N$ is constant, $m$ is restricted to values
$m=-J,...,J$. The unitary operator $e^{\frac{\theta}{2}(a^{\dagger}b-ab^{\dagger})}$ is 
known as the two-mode displacement operator with real displacement parameter $\theta$.

Interestingly, the eigenstate for $m=-J$ corresponds to a coherent state, which gives an 
appropriate description of several physical aspects of the two-mode Bose-Einstein condensate \cite{gordon}.
Its easy to verify that $e^{\frac{-\theta}{2}(a^{\dagger}b-ab^{\dagger})}|J,m\rangle$ 
are the eigenstates of the Hamiltonian (\ref{hamiltonian3}). One must simply apply the two-mode 
displacement operator to the Hamiltonian:
\begin{equation}\label{eq:diagonal}
H_0=A_1(a^{\dagger}a-b^{\dagger}b)+A_2(a^{\dagger}a-b^{\dagger}b)^2+A_3(a^{\dagger}a-b^{\dagger}b)^3,
\end{equation}
which is diagonal in the Fock basis. The result of this transformation is the Hamiltonian $\mathcal{H}_3$
with the coefficients shown in (\ref{eq:coef}), except for an energy shift.
For this reason both Hamiltonians have the same energy spectrum and their eigenvectors are related by the 
displacement operator. The ground state of the system is 
$\ket{E_g}=e^{\frac{-\theta}{2}(a^{\dagger}b-ab^{\dagger})}|J,m_0\rangle$, where $m_0$ is the integer
 that minimizes the energy $E_m=A_1m+A_2m^2+A_3m^3$. This number can be determined with the expressions:
\begin{eqnarray}\label{eqn:min}
m_{0}^{\pm}&=&\frac{A_2}{3\,A_3}\left(-1\pm\sqrt{1-\frac{3A_1A_3}{A_2^2}}\right)\,,\, A_3\neq0\nonumber\\
m_0&=&-\frac{A_1}{2\,A_2}\,,\, \qquad \qquad \qquad \qquad \quad A_3=0.
\end{eqnarray}
If $3A_1A_3/A_2^2>1$, $m_0$ is a complex number and the
energy has no local minimum. Therefore, the minimum energy will
correspond to the extreme point $m_{ex}=-NA_3/|A_3|$. On the other hand,
if $3A_1A_3/A_2^2<1$ then the minimum, which is given by
Eq.(\ref{eqn:min}), is $m_{0}^{+}$ for $A_3>0$ and $m_{0}^{-}$ when
$A_3<0$. However, if the size of the system is big enough then the
cubic part of the energy prevails for $|m|>>0$.
Under this circumstances the energy of the extreme point $E_{m_{ex}}$ is
smaller than the local minimum $E_{m_0^{\pm}}$.

%Since $-N<m<N$, it is possible for
%N to be sufficiently large such that the energy is still minimized at
%$m_0=-\frac{NA_3}{|A_3|}$ and this must be checked.

\begin{figure}[h!]
\begin{center}
\subfigure[]{
\includegraphics[width=0.45\textwidth]{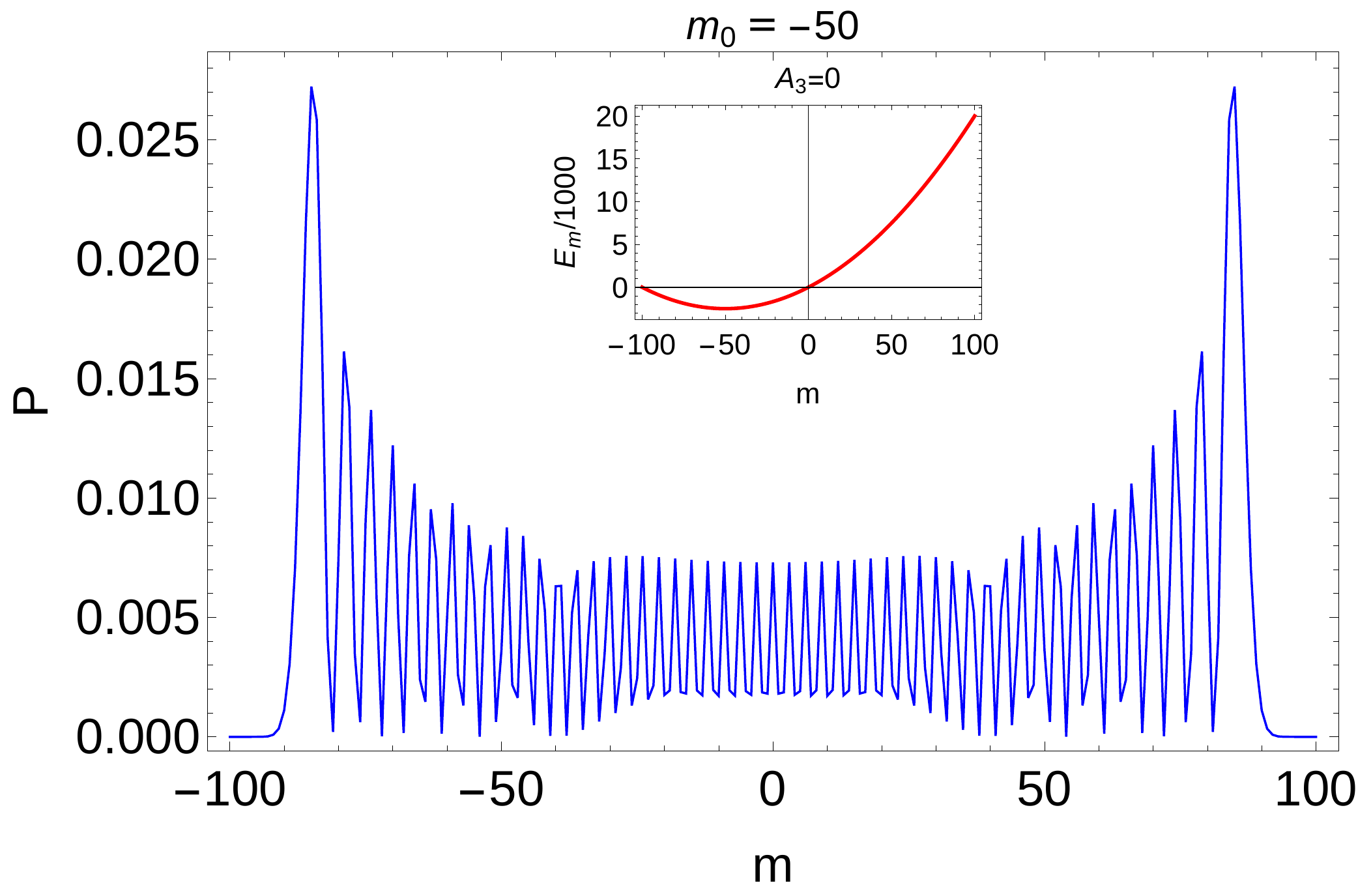}}
\subfigure[]{
\includegraphics[width=0.45\textwidth]{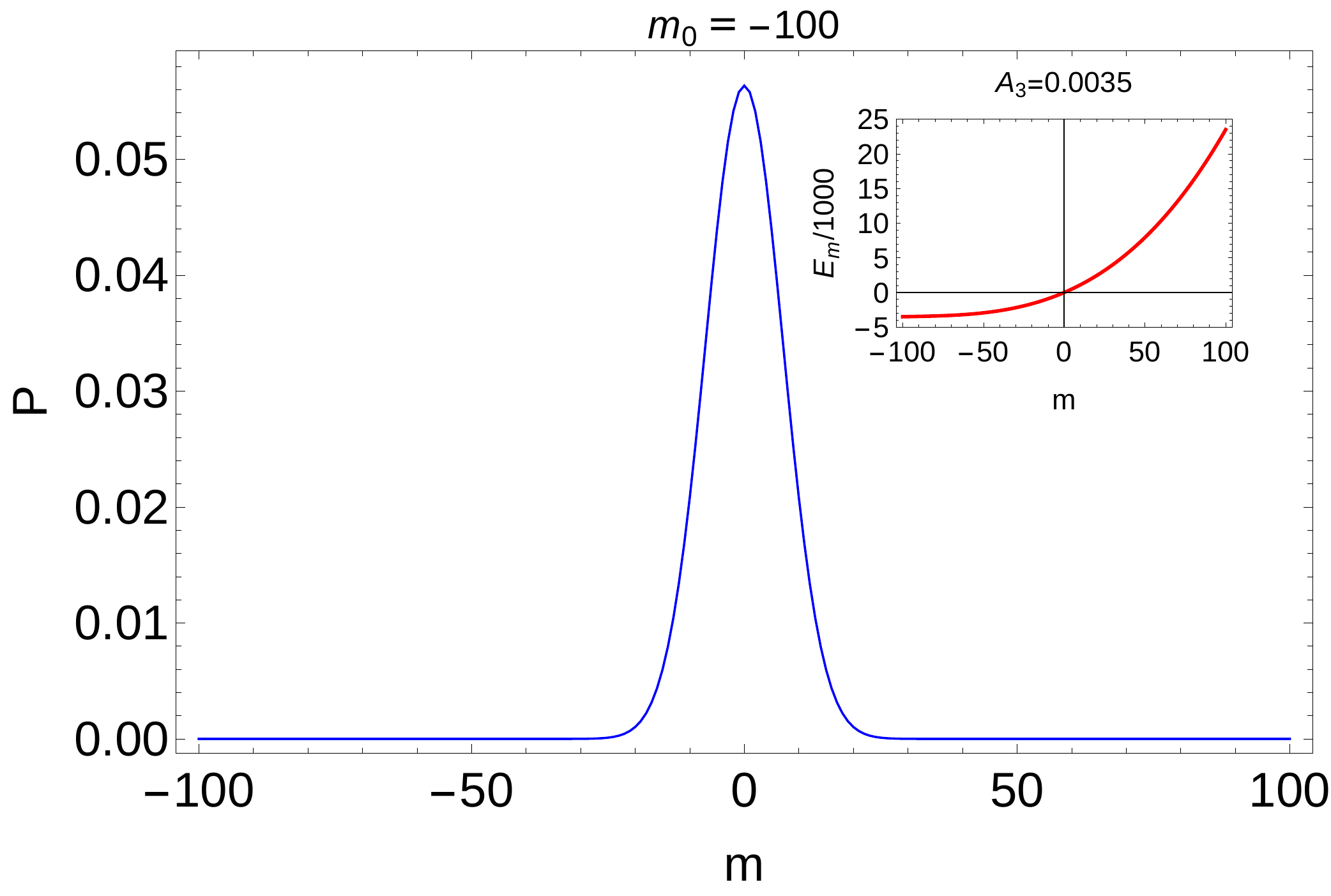}}
\caption{Probability distribution P of the relative population for the ground state with N=100,
$A_1/A_2=100$ and a) $A_3=0$, b) $A_3=0.0035$. The ground state
distribution changes from a multi-peak to a single peak distribution if
three-body collisions are included. The insets show the change in the spectrum.}
\label{figure2}
\end{center}
\end{figure}

A quantity of interest is the probability distribution of the relative population for the ground state, which is given by:
\begin{equation}\label{eq:P}
P=|\langle
N,m|\psi_0\rangle|^2=|d_{m,m_{0}}^{N}|^2 
\end{equation}
where:
\begin{eqnarray}\label{eq:d}
d_{m,m_{0}}^{N}&=\sum_k(-1)^{k-m_0+m}\frac{\sqrt{(N+m_0)!(N-m_0)!(N+m)!(N-m)!}}{(N+m_0-k)!k!(N-k-m)!(k-m_0+m)!}\nonumber\\& 
\large(\cos{\frac{\theta}{2}}\large)^{2N-2k+m_0-m}\large(\sin{\frac{\theta}{2}}\large)^{2k-m_0+m}.
\end{eqnarray} 
are the Wigner rotation matrix elements \cite{sakurai}.
Note that the sum must be done for the values of $k$ such that none of the
arguments of the factorials in the denominator are negative. Different ground
states parametrized by $m_0$ are obtained by changing the rate $A_1 A_3/A^2_2$.
We plot in Fig.(\ref{figure2}) an example for $N=100$ particles with $A_3=0$
(i.e. assuming there are no third order tunneling and three-body collisions)
and $m_0=A_1/2A_2=-50$. Such a state has a multi-peak distribution.  However,
if $A_3=0.0035$ we get a single peak distribution corresponding to $m_0=-100$,
i.e. a coherent state. Each figure has an inset with a plot of the corresponding
spectrum, where we can see that three body processes generate a shift of the minimum.
This shows how three-body collisions and third order
tunneling drastically change the structure of the ground state of the system. Moreover, the energy gap between the ground and first excited states is larger in Fig. (\ref{figure2})b). This is a typical feature of the self-trapping regime \cite{qpt} which suggests that three-body terms tend to favor localization. We will confirm this insight below. 

\begin{figure}
\begin{center}
\subfigure[]{
\includegraphics[width=0.30\textwidth]{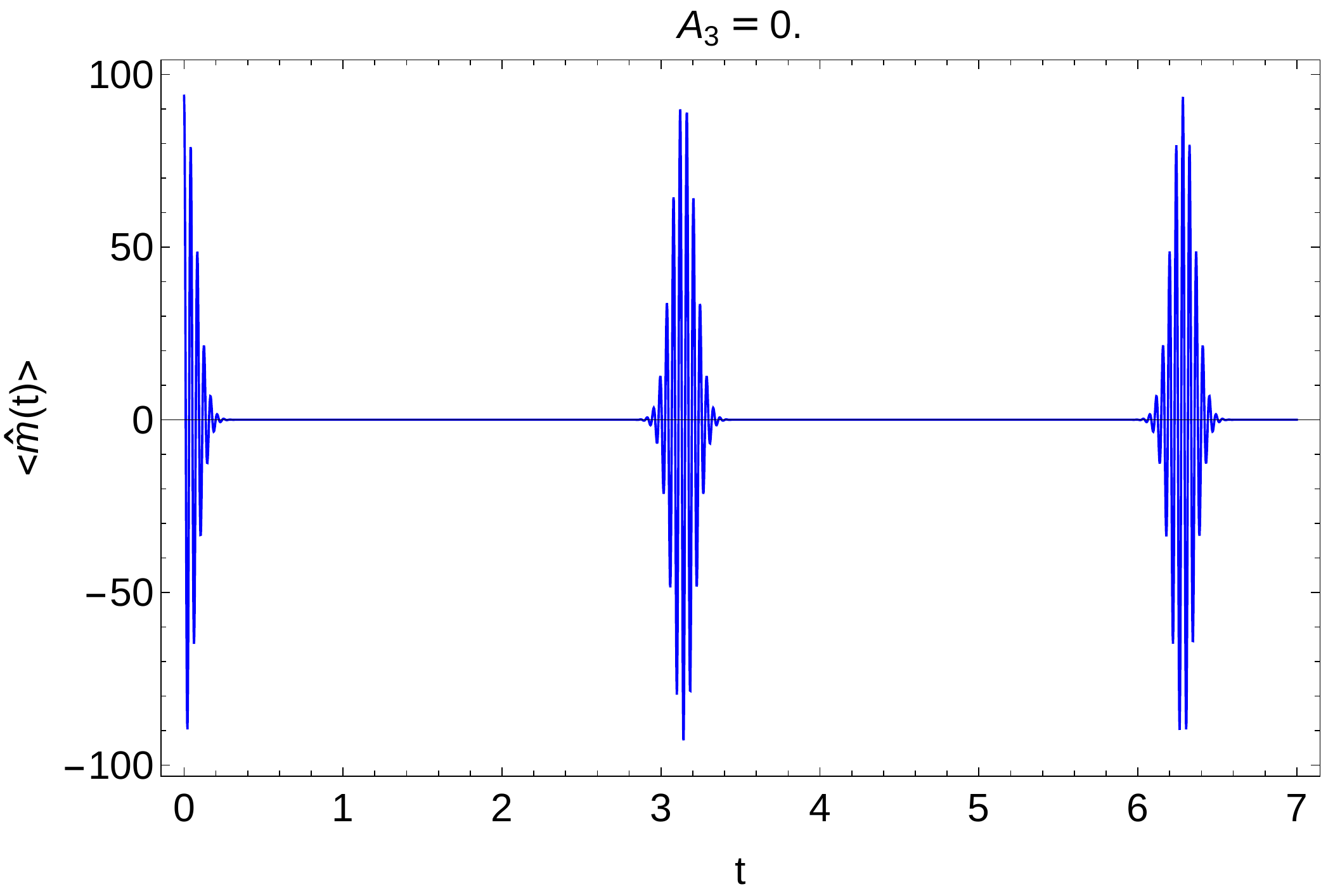}}
\subfigure[]{
\includegraphics[width=0.3\textwidth]{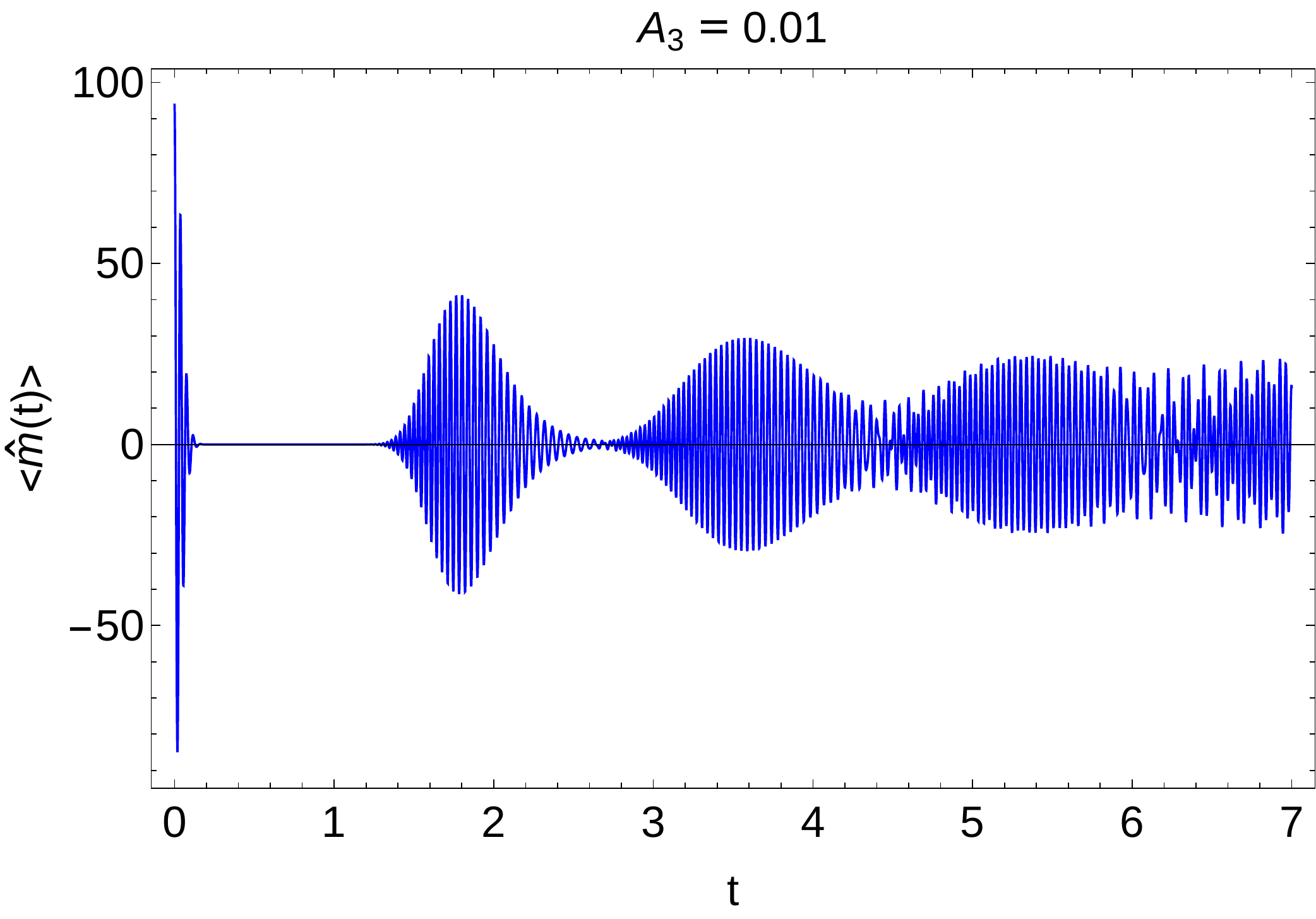}}
\subfigure[]{
\includegraphics[width=0.3\textwidth]{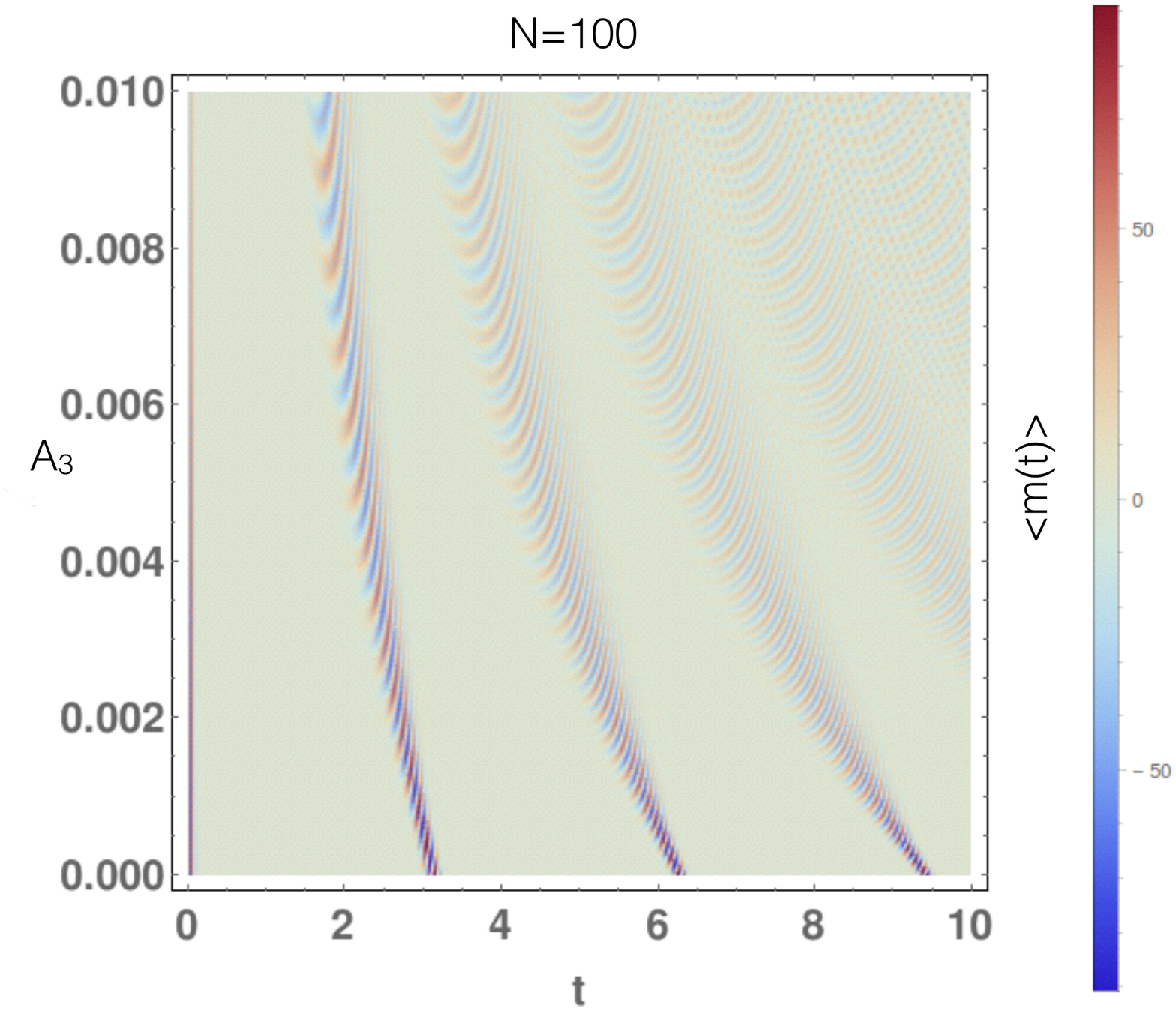}}
\\
\caption{Evolution of the relative population $\langle m\rangle$
for N=100 particles with $A_1=100$, $A_2=1$ and $A_3$ specified in each figure. The initial state corresponds to
$|\psi(t=0)\rangle=|N,N\rangle$. We observe that the presence of three-body collisions changes the dynamics of the system and in fact tends to breakdown perfect quantum collapse-revival cycles. In c), we show that this phenomenon occurs in a continuous fashion with respect to the parameter $A_3$.}
\label{figure3}
\end{center}
\end{figure}

We now analyze the effects of three-body collisions in the evolution of the average relative population
$\langle m\rangle=\langle a^\dagger a-b^\dagger b\rangle$, for an
initial condition $|\psi(t=0)\rangle$. The evolution of the relative
population is given by:
\begin{eqnarray}\label{eq:jzev}
\langle
m\rangle&=&\cos\theta\sum_{-N}^{N}m\,|C_m|^2\\&-&\sin\theta\sum_{-N+1}^{N}
C_{m} C_{m-1}\sqrt{N(N+1)-m(m-1)} L_{m}\nonumber\\
L_{m}&=&\cos\large[(E_{m-1}-E_m)\,t\large]
\end{eqnarray}
where the
coefficients $C_{m}$ are defined by
\begin{equation}\label{eq:cs}
|\psi(t=0)\rangle=\sum_{m=-N}^{N} C_m e^{\frac{-\theta}{2}(a^{\dagger}b-ab^{\dagger})}|N,m\rangle
\end{equation}
In Fig. (\ref{figure3}) we plot equation (\ref{eq:izev}) for $N=100$, $A_1=100$, $A_2=1$ and the initial condition
$|\psi(t=0)\rangle=|N,N\rangle$, i.e. all the particles start in a single well.  We can see that the relative 
population shows collapses and revivals for two body collisions. We can also observe that a small rate of three-body
collisions ($A_3=1/100$) has a noticeable effect on the behaviour of the time evolution of the system, and in fact
it tends to breakdown the perfect collapse-revival cycles of the relative population. These cycles are characteristic of a delocalized dynamics, while the small-amplitude oscillations of the population imbalance in Fig. (\ref{figure3})b) point to a self-trapped dynamics. This confirms that the three-body terms tend to favor localization in this model.

\section{Discussion and conclusions}

At this point, let us discuss the experimental relevance of our results.  In the case of a double-well potential, the modes a and b are quasilocalised modes in wells A and B and $A_1$ corresponds to the asymmetry between the two wells. Thus $\epsilon$ and $A_1$ can be experimentally tuned by changing the distance and the energy offset between the wells, respectively. Indeed, as discussed in \cite{fuent3} the condition $\epsilon << A_1\,\theta$ holds in a wide variety of experiments with double-well BECs \cite{spatially,doublewell1,doublewell2}. 
More specifically, the linear tunnelling rate $ U_{1|2}=\int\,dr\,\phi_1H_t\phi_2 \simeq A_1\theta+\epsilon$ 
takes experimental values ranging from $5\cdot10^{-4} Hz\cdot h$ \cite{doublewell1} to $2\,Hz\cdot h$ 
\cite{spatially}, while the energy offset between the wells can be as high $A_1=530\,Hz\cdot h$ \cite{doublewell2}. Even if the wells are intended to be perfectly symmetric, the uncertainty in the trap depth leads us to assume a minimum trap asymmetry of $A_1\simeq20\,Hz\cdot h$ \cite{spatially}. Regarding the other parameters, $A_2$ and $A_3$ represent the two-body and three-contributions to the potential energy. It has been estimated the latter represent a few percents of the former. For instance, in He \cite{three} $A_3$ is around 2 $\%$ of the total potential energy. All these values are in line with the plots in the previous section. Note also that the three-body terms represent the leading-order corrections to the standard two-body interaction. It is reasonable to assume that four-body and higher order terms will be small with respect to the three-body ones and overall negligible. 
We can conclude that our model is suitable to describe a wide variety of experiments involving double-well BECs. It would be interesting to explore as well the microscopic derivation of a spin-$1/2$ BEC in order to determine the connection with our model.

In summary,  we introduce a model of two-mode Bose-Einstein condensate which includes not only two-body but also three-body interactions. We find an analytical solution and provide the full spectrum of eigenvalues and the corresponding eigenvectors. This allows us to analyse the role of three-body interactions in physical quantities of interest, such as the probability distribution of the relative population or the time evolution of its expectation value. We find that three-body collisions have non-trivial effects, such as significant changes in the probability distribution of the ground state or the inhibition of collapses in the evolution of the relative population of the modes. Our work provides insights on the effects of higher order collisions in the physics of a two component Bose-Einstein condensate. Following the formalism employed in this paper, higher-order collisions can also be included in the model \cite{fuent1}. 

%Interestingly, by including higher order collision, any energy spectrum can be approximated.

\section*{Acknowledgments}
I. F. and C.S acknowledge funding from EPSRC (CAF Grant No. EP/G00496X/2 to I. F.) R.B.M. was supported by the Natural Sciences and Engineering Research Council of Canada.
%\section*{Author contributions}
%IF conceived the main idea and directed the project. IF, CS, P B-B, CH and RBM contributed to the computations, discussion and manuscript preparation.
%\section*{Additional information}
%The authors declare no competing financial interests. Correspondence and requests for materials should be addressed to C. S. (\verb"carlos.sabin@nottingham.ac.uk").

%I. F. was supported by EPSRC and the Alexander von Humboldt
%Foundation and would like to thank Tobias Brandes and his group at
%TU-Berlin for their hospitality.
%

\end{document}